\documentclass[%
 reprint,
 amsmath,amssymb,
 aps,
]{revtex4-1}

\usepackage{graphicx}% Include figure files
\usepackage{dcolumn}% Align table columns on decimal point
\usepackage{bm}% bold math
\usepackage{amssymb} 
\usepackage{braket}
\usepackage{mathcomp}
\usepackage{amsmath}

\begin{document}

%\preprint{APS/123-QED}

\title{Creation of single-photon entangled states around rotating black holes}% Force line breaks with \\

\author{Ovidiu Racorean}
 %\altaffiliation[Also at ]{General Direction of Information Technology, Bucharest, Romania.}%Lines break automatically or can be forced with \\
 \email{ovidiu.racorean@mfinante.ro}
\affiliation{%
General Direction of Information Technology, Bucharest, Romania.\\
}%


%\affiliation{
%\affiliation{
%\affiliation{%


\date{\today}% It is always \today, today,

\begin{abstract}
Recently, numerical simulations showed that X-ray photons emitted by accretion disks acquire rotation of polarization angle and orbital angular momentum due to strong gravitational field in the vicinity of the rotating black holes. Based on these two degrees of freedom we construct a bipartite two-level quantum system of the accretion disk’s photons. To characterize the quantum states of this composite system we consider linear entropy for the reduced density matrix of polarization with the intention to exploit its direct relation with the photons degree of polarization. Accordingly, the minimum degree of polarization of X-ray radiation located in the transition region of the accretion disk indicates a high value of the linear entropy for the photons  emitted on this region, inferring a high degree of entanglement in the composite system. We emphasize that for an extreme rotating black hole in the thermal state, the photons with energies at the thermal peak are maximally entangled in polarization and orbital angular momentum, leading to the creation of all four Bell states. Detection and measurement of quantum information encoded in photons emitted in the accretion disk around rotating black holes may be performed by actual quantum information technology.
\begin{description}
\item[PACS numbers]
% \verb+\pacs{04.70.Dy}+ 04.70.Dy 
 04.70.Dy, 03.67.Bg, 42.50.Ex, 95.30.Gv
\end{description}
\end{abstract}

\pacs{Valid PACS appear here}% PACS, the Physics and Astronomy
\maketitle

%\tableofcontents

\section{Introduction}

Rotating black holes (RBH) are among the most mysterious predictions general relativity has made. Spinning black holes are literally curving and twisting the space-time nearby influencing the photons emitted by the accretion disk. The strong gravitational field in the vicinity of spinning BH rotates the angle of polarization and imprint orbital angular momentum (OAM) to X-ray photons emitted by the accretion disk. 

The efforts to determine the hidden characteristics of rotating black holes, such as the speed of spinning, were mainly focused in the recent years over detecting the polarization of X-ray radiation coming from the accretion disk. It was pointed out in \cite{sch}, \cite{chan}, \cite{con}, \cite{conn}, \cite{cun}, \cite{agol}, based on Stokes parameters calculation, that photons emitted by the accretion disk of black holes in the thermal state, should possess linear polarization, either parallel or perpendicular to the plane of the disk. Thus, on the outer region of the disk, at low energies, the X-ray radiation is horizontally polarized parallel to the plane of the disk. The polarization angle of photons coming from the innermost region is shifted through vertical polarization, perpendicular to the disk plane due to the strong gravitational field in the vicinity of the RBH. The degree of polarization is higher in these two regions of the accretion disk. Between the outer and inner regions of the disk, at the transition region, the relative contributions of horizontal and vertical polarized photons are nearly equal and no net polarization is observed. The transition region is characterized by very low degrees of polarization that tends to zero for extreme spinning black holes.

On the other hand, numerical simulations \cite{tam}, \cite{yang}, \cite{eli}, \cite{uri}, suggest that strong dragging frame effect near rotating black holes imprint nontrivial orbital angular momentum (OAM) modes to photons emitted by the accretion disk. X-ray photons coming from the disk are forced by the twisted spacetime around spinning black holes to acquire OAM. As the BH spins faster the OAM spectrum extend to wider values carrying both positive and negative OAM ($\pm\ell$). 

When measured by a distant observer, the X-ray radiation coming from accretion disk should carry linear polarization (horizontal or vertical) and specific (positive or negative) OAM values as consequence of strong general relativistic effects that occur in the presence of the rotating black holes. We further develop the analysis of radiation emitted near RBH considering here the polarization and OAM modes as two degrees of freedom that can fully specify the states of the accretion disk's X-ray photons.

Generically, the states of X-ray photons emitted near rotating black holes should be determined by constructing the 4x4 density matrix \cite{mcm1}, \cite{mcm2}, of the composite system consisting of polarization and OAM modes, as two distinct subsystems. In order to avoid the complications that arise in constructing this bipartite system density matrix we focus our attention over the reduced density matrix of polarization subsystem, which is easy inferred from Stokes parameters. Our main assumption here is that the degree of mixedness of one subsystem determines the degree of entanglement of the composite system, \cite{gam}, \cite{chi}, \cite{hor}, \cite{bjo}, \cite{gen}, \cite{wei}. Moreover, a high degree of mixedness present in the polarization subsystem suggests a high degree of non-separability (entanglement) of the composite system.

We consider the linear entropy of the X-ray photons polarization in order to shed light on the degree of mixedness in this subsystem. The decision to chose here the linear entropy as the measure of the degree of mixeedness of the polarization subsystem was inspired by its relation to the degree of polarization.

The linear entropy values vary from low levels in the outer and the innermost regions of accretion disk of rotating black holes, regions characterized by a high degree of polarization, to very high levels at the transition region where a very low degree of polarization is encountered. Based on the linear entropy values we conclude that at the composite system level, the photons at the outer and innermost regions of the disk are characterized by high degrees of separability, while the X-ray originated in the transition region are endowed with a high degree of nonseparability (entanglement). 

The degree of entanglement of X-ray photons in the transition region is influenced (via the degree of polarization) by the speed of BH rotation. Broadly, the faster the BH spins the higher the degree of entanglement of X-ray photons in polarization and OAM. Under these circumstances, it is expected that the maximally entangled states to appear in the case of an extreme spinning BH ($a=1$), for the photons at the energy peak in the transition region.

The maximal entanglement in polarization and OAM of X-ray photons coming from the transition region of the accretion disk is expressed via Schmidt decomposition \cite{ber}, \cite{eke}, by the all four Bell states.

The entangled states of X-ray photons can be measured by quantum optics setups \cite{kho}, \cite{gun}, with the limitation imposed here by the high energy of photons (X-ray band). With the later progress in technology related to X-ray laboratory research \cite{sas}, \cite{pel}, \cite{kho}, leading to quantum computation with X-ray photons \cite{per} legitimate hopes are directed towards detection and measurements of the Bell states the X-ray radiation coming from galactic active nuclei and solar-mass black holes accretion disk.

Detection of Bell states is an important indication of the possibility that strong gravitational field near rotating black holes to literally implement complex quantum information processes encoded in the X-ray photons emitted by the accretion disk.

\section{X-ray polarization from accreting black holes}

Expectations to unveil some of the black holes hidden characteristics, such as the speed of spinning, are related to measuring the polarization of X-ray photons emitted by the accretion disk. In analysing the polarization of radiation emitted by the accretion disk in thermal state we consider the simplest model of a geometrically thin, optically thick, steady-state accretion disk, aligned with the BH spin axis. 

Stokes parameters are central in attempts \cite{sch}, \cite{chan}, \cite{con}, \cite{conn}, \cite{cun}, \cite{agol}, to estimate the degree of polarization and the polarization angle of X-ray radiation emitted by accretion disk, the two parameters that quotes the characteristics of black holes. Considering that the polarization of X-ray photons is induced by Compton scattering, which prevent the circular polarization, the Stokes parameters are:  

\begin{equation}
s_0=I
\end{equation}                               
\begin{equation}
s_1=Q=I cos2\chi  
\end{equation}
\begin{equation}
s_2=U=I sin2\chi,  
\end{equation}

where $\chi$ is the polarization angle.

The angle of polarization and the degree of polarization, are derived from the Stokes parameter in the following manner :

\begin{equation}
tan2\chi=\frac {U}{Q}  
\end{equation}
and
\begin{equation}
\delta=\sqrt{Q^2+U^2}  ,
\end{equation}
respectively.

These two parameters are sensitive to the black hole spin ($a$) and the inclination angle ($i$) of the accretion disk in relation to a distant observer that detect X-ray photons coming from the accretion disk. The angle of inclination is not relevant here and will be assigned to a constant value  $i=〖45〗^0$ ; only the spin parameter of the black hole is inferred further.

Recent numerical simulations \cite{sch}, scaled up the dependency of the degree of polarization and the angle of polarization by the spin parameter of rotating black holes. It was shown that X-rays emitted near the outer region of the accretion disk, at low energies, are horizontal ($H$) polarized, parallel to the plane of the disk, with the angle of polarization $\chi=0$. The strong gravitational field close to the black hole rotates the polarization angle of the X-ray photons emitted by the innermost regions of the disk to positive or negative values of vertical ($V$) polarization, perpendicular to the plane of the disk, with the polarization angle $\chi=\pm90^o$.

The prescription for the influences of the BH's spin parameter ($a$) over the  degree of polarization of X-ray radiation are resumed in the Fig. 1.   

\begin{figure}
\includegraphics[width=8.6cm]{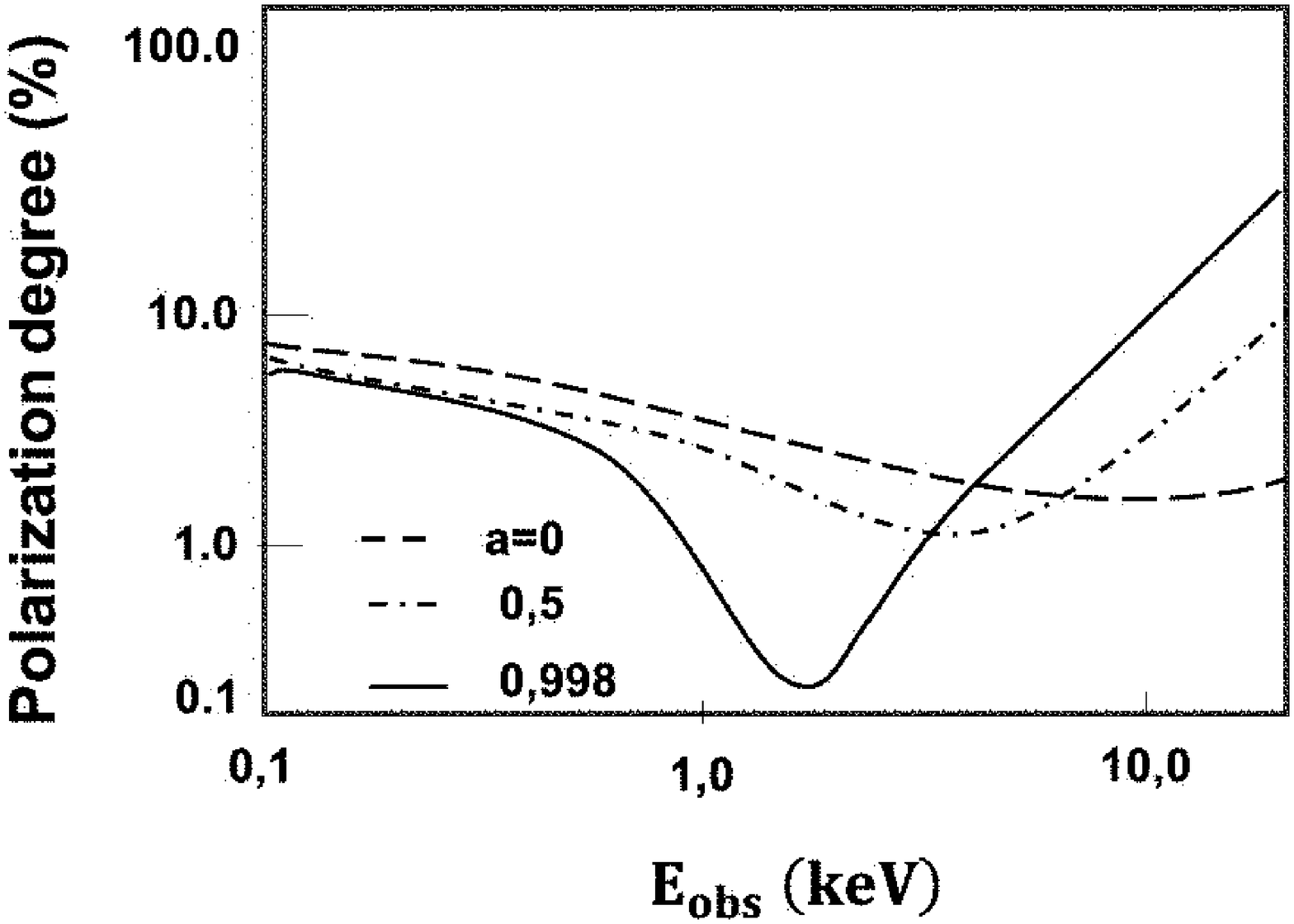}
\caption{\label{fig:DOP} The degree of polarization of X-ray photons emitted by accretion disks of black holes having different spin parameter}
\end{figure}

High levels of degree of polarization could be noticed in Fig.1 for the outer and innermost regions of the accretion disk, regions characterized by strong horizontal and vertical polarization.  

At the transition region of the disk, situated between low and high energies of X-ray radiation, at energies peaking around 1 keV, both horizontal and vertical polarization photons are present canceling each other contribution, phenomenon reflected in a minimum of the degree of polarization. 

It can be noted from the Fig.1 that for near-extreme rotating black holes ($a=0,998$) the degree of polarization of photons at the peak energies emitted in the transition regions goes to a minimum that almost approaches zero value, $\delta\approx0$. 

The degree of polarization reaches zero value, $\delta=0$, for extreme spinning black holes ($a=1$). 

Although, the angle of polarization and the degree of polarization provide important information about the physical medium near RBH, we invoke here an alternative and less considered approach \cite{mcm1}, \cite{mcm2}, that promises to reveal new aspects related to radiation emitted by accretion disks. This approach infers the Stokes parameters to construct the polarization matrix of the photons emitted near rotating black holes; polarization matrix unitary contains all physical information about the polarization state of radiation.

The polarization matrix considering only the linear polarization is given by: 

\begin{equation}
    \rho_{pol} = \frac{1}{2}\begin{bmatrix}I+Q & U  \\U & I-Q \end{bmatrix}
\end{equation}

Replacing the values of Stokes parameters from the Eq. (1), polarization matrix takes the form:

\begin{equation}
    \rho_{pol} = \begin{bmatrix}cos^2\chi & cos\chi  sin\chi \\cos\chi  sin\chi& sin^2\chi \end{bmatrix}
\end{equation}

The polarization matrix equals the density matrix of the photons source \cite{gam}, \cite{chi}, since the density matrix is the unit trace scaling of the polarization matrix and $Tr(\rho_{pol} )=1$, such as for the rest of the present paper we will refer to $\rho_{pol}$ as the density matrix of polarization of X-ray photons.

The matrix in Eq. (7) is the Hermitian density matrix of the two-level orthogonal system of X-ray photons polarization. Throughout the density matrix representation, polarization could be considered as a X-ray photon degree of freedom that could encode quantum information. 

The states of the X-ray photons are included in the two dimensional Hilbert ($H_{pol}$) space of polarization $\rho_{pol} \in H_{pol}$.
 
The density matrix (7) corresponds to the general quantum state of X-ray photons emitted by the accretion disk near the rotating black hole: 

\begin{equation}
\ket{\Phi_{pol}}=cos\chi \ket{H}+sin\chi \ket{V}
\end{equation}

where $H$ and $V$ are horizontal and vertical polarization.

\section{Orbital angular momentum of photons near rotating black holes}

Acquiring orbital angular momentum by photons emitted or passing nearby the equatorial plane of rotating black holes, the hypothesis formulated by Harwit \cite{har}, gained consistency lately in \cite{tam}, \cite{yang}, \cite{eli}, \cite{uri}. It is suggested that the strong frame dragging effect caused by the spinning black holes may force radiation beams to acquire OAM modes.  

In support to this idea, Tamburini et al. \cite{tam}, performed numerical simulations of the radiation emitted by the accretion disk of rotating black holes and observed the generation of nontrivial photon OAM modes that form asymmetric spectra in terms of the LG-modes. Radiation beams emitted by the accretion disk acquiring independent azimuthal phase term $e^{i\ell\phi}$ possess an orbital angular momentum of $\ell$ per photon, where $\ell$ is the integer topological charge. It was pointed out that the initially zero OAM mode of photons flips to wider OAM modes is independent of the photon frequency, such as we choose to refer in what follows to X-ray radiation. 

The OAM modes $\ell$ that the frame dragging effect nearby spinning black holes imprint to X-ray radiation emitted by accretion disk are determined by the BH spin parameter ($a$) and the inclination angle of the disk ($i$) towards a remote observer. We maintain for the inclination angle the same value $i=〖45〗^0$ as discussed in the case of X-ray polarization and analyze further only the BH spin parameter influences on the radiation coming from the accretion disk.

The Fig.2 depicts the spectrum of OAM modes acquired by X-ray photons emitted near RBH that have two different values for the spin parameter, a relatively moderate spinning BH ($a=0,5$) to the left and a near-extreme rotating BH ($a=0,99$) to the right. 

\begin{figure}
\includegraphics[width=8.6cm]{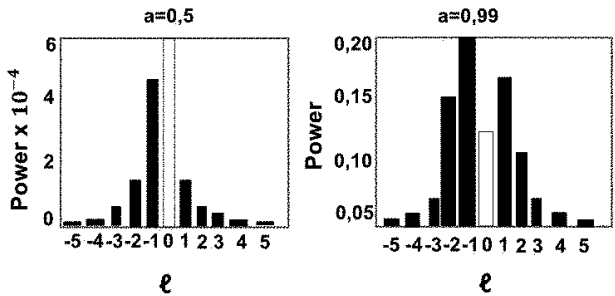}
\caption{\label{fig:OAM} OAM modes spectrum of X-ray photons emitted by the accretion disk of two different spin parameter black holes:  slow spinning BH (a) and fast spinning BH (b)}
\end{figure}

Notice from Fig.2 that the moderate spinning BH is characterized by a narrow spectrum of OAM modes that reduces to zero LG mode ($\ell=0$) for Schwarzschild BH ($a=0$). The zero OAM mode is still dominant for moderate rotating BH with the spin parameter, $a=0,5$. 

A rapidly spinning black hole ($a=0,99$) has the innermost region of the accretion disk closer to the black hole event horizon. X-ray photons emitted by the accretion disk are influenced stronger by the gravitational field because of the BH proximal vicinity and tends to flip toward wider OAM modes. As it can be noticed from Fig.2 the near-extreme spinning Black hole has a broader OAM spectrum with a significant reduction of zero LG mode.

A remarkable aspect that reside from the OAM maps above is that X-ray photons emitted near the RBH could acquire both negative and positive OAM values, $-\ell$ and $\ell$ . Our interest in this particular aspect is motivated by the fact that negative and positive OAM values can be viewed as a degree of freedom of photons, which could encode quantum information.

The photons OAM modes ($\pm\ell$) are orthogonal \cite{mai}, \cite{kre}, and form an unbound Hilbert space, unlike the case of photon polarization which is two-dimensional. The OAM modes ($\pm\ell$) can possess theoretically any values in the interval $[-\infty,\infty]$. We restrict the OAM unbound Hilbert space to a two dimensional Hilbert($H_{OAM}$) space by considering a generic subspace of OAM ($\pm\ell$), were $\ell$ can take precise values $\pm1, \pm2,... $ , as it will be seen later. For now we will consider generic OAM modes ($\pm\ell$) having in mind a specific value of X-ray photons OAM.

The generic X-ray photon emitted nearby RBH acquires not only a definite polarization (horizontal or vertical) but also a definite value of OAM ($\pm\ell$). To consistently specify the state of X-ray photons emitted by the accretion disk both of the two degrees of freedom must be considered. The polarized photon has now one more degree of freedom to assess its quantum state, the OAM modes, such as a photon horizontally/vertically ($H$/$V$) polarized also possess $-\ell$ or $\ell$ OAM modes. 

The states of photons emitted by the RBH’s accretion disk have to be expressed by a 4x4 density matrix that corresponds to a bipartite quantum system:

\begin{equation}
\rho_{pol-OAM} \in H_{pol}\otimes H_{OAM}
\end{equation}

The states of X-ray photons emitted near RBH defined by a bipartite system, polarization – OAM have the general form:

\begin{equation}
\ket{\Phi_{pol-OAM}}=\alpha_{00}\ket{H,\ell}+\alpha_{01}\ket{H,-\ell}+\alpha_{10}\ket{V,\ell}+\alpha_{11}\ket{V,-\ell}
\end{equation}

with $\sum_{ij} \alpha_{ij}^2=1$.

\section{Linear entropy of accreting black holes photons}

A good measure to characterize the state of the accretion disk photons having acquired the two degrees of freedom (polarization and OAM) is the linear entropy of the system. The linear entropy expresses the degree of mixedness of a composite system and is related to the density matrix through the relation:

\begin{equation}
S_L=\frac{d} {d-1} ( 1-Tr(\rho^2))
\end{equation}

where $d$ is the dimension of the Hilbert space of the system, here $d=4$ for the X-ray photons with two degrees of freedom, and $Tr$ stands for the trace of the density matrix of the composite system. 

In order to determine the values of the linear entropy that specify the state of the composite system one should construct the 4x4 density matrix, a process that is  notoriously difficult. To avoid the difficulties related to the construction of the density matrix of the accretion disk photons, we choose here to follow a different path. We consider to determine the state of the composite system by exploring the states of its subsystems - polarization and OAM modes.  
Our main assumption is that the state of a subsystem could be inferred to accurately specify the state of the composite system, \cite{gam}, \cite{chi},. Accordingly, a high degree of mixedness present in one subsystem implies a high degree of nonseparability (entanglement) of the composite system and vice-versa, a large amount of purity in the subsystem characterize a composite system in the mixed state. 

Benefiting from the fact that polarization subsystem is to a certain extent known, we consider here the reduced density matrix of the polarization subsystem in order to determine the state of the composite system, polarization-OAM.

The reduced density matrix of polarization of the bipartite -polarization-OAM- composite system is found by taking a partial trace over the OAM states: 

\begin{equation}
\rho_{red}= Tr_{OAM}(\ket{\Phi_{pol-OAM}}\bra{\Phi_{pol-OAM}})
\end{equation}

where $Tr_{OAM}$ is the trace over OAM modes of the density matrix on the composite system, the X-ray photon having two degree of freedom – polarization and OAM.

The reduced density matrix, in our case is the polarization matrix in Eq. (7) since two degrees of freedom that compose the bipartite system polarization-OAM refers the same photons.

The linear entropy for the X-ray polarization subsystem considering the reduced density matrix can be written as: 

\begin{equation}
S_L=2 ( 1-Tr(\rho_{pol}^2))
\end{equation}

Noticing here that there is a complete equivalence between quantum purity ($Tr(\rho_{pol}^2)$)and degree of polarization for single photons we can write:

\begin{equation}
Tr(\rho_{pol}^2)=\frac{1}{2}(1+\delta^2)
\end{equation}

The linear entropy for the X-ray photons polarization subsystem takes the simple form \cite{chi}:

\begin{equation}
S_L= 1-\delta^2
\end{equation}

To determine the values of the linear entropy and infer over the states of the composite system we should refer to the degree of polarization in Fig.1. We observe high degree of polarization of the photons emitted in the outer and the innermost regions of the accretion disk which induces, according to Eq.(15), low values of linear entropy, almost close to zero, $S_L\approx0$. The photons emitted in these two regions of the disk, characterized by strong horizontal and vertical polarization, can be considered as being closer to a pure state. 
The high degree of purity found in the polarization subsystem indicates a composite system in a state of mixedness.

The minimum level of the degree of polarization, inferred from Fig.1, is observed for the X-ray photons in the transition region of the accretion disk , at the thermal peak of the energy. We conclude that at the thermal peak the linear entropy approaches very high values, $S_L\approx1$, which induces a state of high degree of mixedness in the polarization of photons at the transition region of the accretion disk. The high degree of mixedness of the photons at the thermal peak determines a high degree of non-separability (entanglement) for the composite system of polarization- OAM considered. The X-ray photons emitted by the RBH accretion disk at the thermal peak are entangled to some extent in polarization and OAM modes. 

An important note we also emphasize from the Fig.1. is the consideration that the faster block hole is spinning the lower the degree of polarization at the thermal peak.  

For extreme spinning black holes the linear entropy reaches its maximum level,$S_L\approx1$, which induces a state of maximal entanglement for the composite system - polarization and OAM. 

\section{Entangled states of photons around rotating black holes}

The high degree of nonseparability of the composite system and the orthogonality of the two-level subsystems suggest that we could further simplify our attempt to explicitly specify the quantum state of X-ray photons emitted by the accretion disk, by performing a Schmidt decomposition.

Considering for the Schmidt decomposition the basis, ${H,V}$ for polarization, and ${\ell,-\ell}$, for the OAM modes, the general state of the composite system in Eq. (10) reduces to the elegant form:

\begin{equation}
\ket{\Phi_{pol-OAM}}=\sqrt{\lambda}\ket{H}\ket{\ell}+\sqrt{1-\lambda}\ket{V}\ket{-\ell}
\end{equation}

where $\lambda$ and $1-\lambda$ are the Schmidt coefficients. The Schmidt coefficients are the eigenvalues of the reduced density matrix of polarization, $\lambda=\cos^2\chi$, and  $1-\lambda=\sin^2\chi$. 

With the Schmidt coefficients determined, the general state of polarization-OAM for the X-ray photons emitted by the accretion disk takes the form:

\begin{equation}
\ket{\Phi_{pol-OAM}}=\cos\chi\ket{H}\ket{\ell}+\sin\chi\ket{V}\ket{-\ell}
\end{equation}

The existence of the two nonzero values of Schmidt coefficients induces ambiguities in determining which state of polarization ($H/V$) should be paired with which state of OAM ($\ell/-\ell$). Consequently, we could consider that the states:

\begin{equation}
\ket{\Psi_{pol-OAM}}=\cos\chi\ket{H}\ket{-\ell}+\sin\chi\ket{V}\ket{\ell}
\end{equation}

must also be present.

Recall that vertical polarization could posses both positive and negative values $\chi=\pm90^o$, such as, other two possible states must not be excluded as possible outcomes of measurements of X-ray photons. These two other states refer to the negative angle of vertical polarization.

All four possible states are synthesized as:

\begin{equation}
\ket{\Phi_{pol-OAM}^\pm}=\cos\chi\ket{H}\ket{\ell}\pm\sin\chi\ket{V}\ket{-\ell}
\end{equation}

and 

\begin{equation}
\ket{\Psi_{pol-OAM}^\pm}=\cos\chi\ket{H}\ket{-\ell}\pm\sin\chi\ket{V}\ket{\ell}
\end{equation}

The Eq.(19) and Eq.(20) express the nonmaximally entangled states of the bipartite system consisting of photons with polarization – OAM degrees of freedom. It should be noticed that the degree of entanglement is determined by the polarization angle value.

A particularly interesting case is reflected by extreme rotating black holes. In the transition region near the extreme spinning black holes the degree of polarization tends to zero, fact that induces a maximally entangled state aver the composite system, polarization -OAM modes. The Schmidt coefficients in this particular case are easy to be calculate as $\lambda=\frac{1}{2}$ and $1-\lambda=\frac{1}{2}$.  

The states of X-ray photons considering the above values for the Schmidt coefficients yields:

\begin{equation}
\ket{\Phi_{pol-OAM}^\pm}=\frac{1}{\sqrt{2}}(\ket{H}\ket{\ell}\pm\ket{V}\ket{-\ell})
\end{equation}

and

\begin{equation}
\ket{\Psi_{pol-OAM}^\pm}=\frac{1}{\sqrt{2}}(\ket{H}\ket{-\ell}\pm\ket{V}\ket{\ell})
\end{equation}

and are the well known Bell states, that reflect the maximally entangled states of X-ray photons in polarization and OAM. 

\section{Discussion}
Although it may seem premature to speak of the detection and measurement of entangled states since neither polarization, nor OAM modes of X-ray radiation emitted near RBH have not been properly observed to date. The tentative detection of X-ray polarization mission GEMS (Gravity and Extreme Magnetism Small Explorer) that NASA has been scheduled to be launched in 2012 was canceled on budget ground.   

The detection and measurement of entangled states are very common these days in optical quantum information processes. Precise setups to measure all four Bell states of single-photons, in polarization and OAM modes, have been reported \cite{kho},\cite{gun},. 

The same technical setups could be inferred in the measurement of X-ray photons considering the limitations determined by their high energies. However, the technology required in detection of the X-ray Bell states had already been tested in the laboratory. 

The later improvement in technology applied to X-ray analysis in the laboratory, such as spiral phase plates, wave plates, beam splitters \cite{pel}, \cite{shw}, and leading to quantum computation with X-ray photons \cite{per}, were performed well in the laboratory tests. Nevertheless, the idea that once detected by telescopes, the X-ray photons emitted at the thermal peak near rotating black holes could be analyzed and the four (nonmaximally) Bell states actually could be measured may not be that exaggerated. 

We stated earlier that to detect entangled states of X-ray photons from the accretion disk, we have to specify the exact two-dimensional Hilbert space of OAM modes. The most probable OAM modes X-ray photons can acquire, as it can be inferred from the figure 2 are $ \ell=\pm1$ , since in the OAM map these two values are the most representative. 

Choosing $\ell=\pm1$ as the basis for the OAM modes the X-ray photons Bell states take the form: 

\begin{equation}
\ket{\Phi_{pol-OAM}^\pm}=\cos\chi\ket{H}\ket{1}\pm\sin\chi\ket{V}\ket{-1}
\end{equation}
and
\begin{equation}
\ket{\Psi_{pol-OAM}^\pm}=\cos\chi\ket{H}\ket{-1}\pm\sin\chi\ket{V}\ket{1}
\end{equation}

The setup destined to measure the Bell states of X-ray photons coming from RBH should be calibrated over the states in Eq.(23)-(24). 

It is expected that X-ray radiation coming from accreting spinning black holes to carry a spectrum of degree of entangled states, from separable states, to nonseparable states and maximally entangled states.

\section{Conclusions}
In conclusion, we argued in the present paper that photons of X-ray radiation emitted by the accretion disk of rotating black holes are entangled in polarization and OAM. The degree of entanglement depends on the region of the accretion disk the X-ray radiation is coming from and the speed BH is spinning. 

Photons emitted by accretion disks, influenced by strong gravitational field nearby spinning black holes acquire OAM and suffers rotation of polarization angle. We consider these two degrees of freedom of photons as a bipartite two-level system and infer that the reduced matrix of polarization determined by the degree of polarization is a measure of the mixedness of the system via linear entropy.

We have shown, based on the high degree of polarization values that X-ray photons emitted in the outer and the innermost regions of the accretion disk in near pure states are very weakly entangled. Radiation emitted in the transition region of the disk probe a weak degree of polarization signaling a higher degree of entanglement in polarization and OAM at the composite system level. 

The degree of polarization in the transition region weakens as the BH spins faster, fact that allows us to conclude that X-ray photons probe higher degrees of entanglement near faster spinning BH. The X-ray photons maximal entangled states are present for extreme rotating BH.

The maximal entanglement is represented by all four Bell states, that we deduced via Schmidt decomposition.

Although it may seem premature to speak on measurement of this Bell states, the present development of quantum information apparatus allows us to hope that near future will probe the entanglement in polarization and OAM for X-ray photons emitted near spinning BH.

 Detection of Bell states in photons coming from RBH may prove to be important since it suggest that these mysterious astrophysical bodies are capable to implement complex quantum information processes.

\end{document}